\documentclass{PoS}

\usepackage{wrapfig}
\usepackage{amsmath}
\usepackage{amsfonts,amsbsy,amsxtra}
\usepackage{amssymb,latexsym}
\usepackage{bm}
\usepackage{textcomp}
\usepackage{ifthen}
\usepackage{lineno}
\usepackage{blindtext}

\newcommand{\figref}[1]{{Fig.~\ref{fig:#1}}}
\newcommand{\Figref}[1]{{Figure~\ref{fig:#1}}}

\newcommand{\equref}[1]{{Eq.~\eqref{eq:#1}}}

\newcommand{\Tabref}[1]{{Table~\ref{tab:#1}}}

\newcommand{\others}{\textit{et al.}}

\newcommand{\antibar}[1]{%
  \overline{#1}%
}
\newcommand{\PPh}{\ensuremath{\gamma}}

\renewcommand{\Pr}{\ensuremath{\rho}}
\newcommand{\Prthree}{\ensuremath{\rho_3}}
\newcommand{\Prz}{\ensuremath{\rho^0}}
\newcommand{\Paone}{\ensuremath{a_1}}
\newcommand{\Patwo}{\ensuremath{a_2}}
\newcommand{\Pafour}{\ensuremath{a_4}}

\newcommand{\Pfzero}{\ensuremath{f_0}}
\newcommand{\Pfone}{\ensuremath{f_1}}
\newcommand{\Pftwo}{\ensuremath{f_2}}

\newcommand{\Po}{\ensuremath{\omega}}
\newcommand{\Ppi}{\ensuremath{\pi}}
\newcommand{\Ppione}{\ensuremath{\pi_1}}
\newcommand{\Ppitwo}{\ensuremath{\pi_2}}
\newcommand{\Ppiz}{\ensuremath{\pi^0}}
\newcommand{\Ppip}{\ensuremath{\pi^+}}
\newcommand{\Ppim}{\ensuremath{\pi^-}}
\newcommand{\Ppipm}{\ensuremath{\pi^\pm}}
\newcommand{\Ppimp}{\ensuremath{\pi^\mp}}
\newcommand{\Peta}{\ensuremath{\eta}}
\newcommand{\Petaprime}{\ensuremath{\eta'}}
\newcommand{\PK}{\ensuremath{K}}
\newcommand{\PKbar}{\ensuremath{\antibar{K}}}

\newcommand{\PKs}{\ensuremath{K^0_S}}
\newcommand{\PKsbar}{\ensuremath{\antibar{K}^0_S}}

\newcommand{\PKp}{\ensuremath{K^+}}
\newcommand{\PKm}{\ensuremath{K^-}}

\newcommand{\Pqq}{\ensuremath{q}}
\newcommand{\Pqqbar}{\ensuremath{\antibar{q}}}
\newcommand{\Pqu}{\ensuremath{u}}
\newcommand{\Pqubar}{\ensuremath{\antibar{u}}}
\newcommand{\Pqd}{\ensuremath{d}}
\newcommand{\Pqdbar}{\ensuremath{\antibar{d}}}
\newcommand{\Pqs}{\ensuremath{s}}
\newcommand{\Pqsbar}{\ensuremath{\antibar{s}}}
\newcommand{\Pqn}{\ensuremath{n}}

\newcommand{\Pp}{\ensuremath{p}}
\newcommand{\Ppbar}{\ensuremath{\antibar{p}}}

\newcommand{\twopion}{\ensuremath{\Ppip\Ppim}}
\newcommand{\threepion}{\ensuremath{\Ppim\Ppip\Ppim}}
\newcommand{\fourpion}{\ensuremath{\Ppip\Ppim\Ppip\Ppim}}
\newcommand{\fivepion}{\ensuremath{\Ppim\Ppip\Ppim\Ppip\Ppim}}

\newcommand{\lrBrk}[1]{{\left[{#1}\right]}}

\newcommand{\abs}[1]{{|{#1}|}}

\newcommand{\measresult}[4]{%
  \ensuremath{#1%
    \ifthenelse{\equal{#2}{}}%
    {}%
    {\pm #2%
      \ifthenelse{\equal{#3}{}}%
      {}%
      {_\text{stat.}}%
    }%
    \ifthenelse{\equal{#3}{}}%
    {}%
    {\pm #3_\text{syst.}}\text{#4}%
  }%
}

\newcommand{\jpc}{\ensuremath{J^{PC}}}

\newcommand{\wavespec}[7]{%
  \ensuremath{#1^{#2#3}\,\allowbreak
    #4^{#5}\,\allowbreak
    \ifthenelse{\equal{#6}{}}%
    {}%
    {[{#6}]}
    #7%
  }%
}

\newcommand{\gevc}{~\ensuremath{\text{GeV}\! / c}}
\newcommand{\gevcsq}{~\ensuremath{(\text{GeV}\! / c)^2}}
\newcommand{\mevcc}{~\ensuremath{\text{MeV}\! / c^2}}
\newcommand{\gevcc}{~\ensuremath{\text{GeV}\! / c^2}}
\newcommand{\tenpow}[2][]{%
  \ifthenelse{\equal{#1}{}}
  {\ensuremath{10^{#2}}}
  {\ensuremath{{#1} \cdot 10^{#2}}}
}

\title{Hadron Spectroscopy in COMPASS}

\ShortTitle{Hadron Spectroscopy in COMPASS}

\author{%
  \speaker{Boris Grube}%
  \thanks{This work is supported by the German BMBF, the
    Maier-Leibnitz-Labor der LMU und TU M\"unchen, the DFG Cluster of
    Excellence \emph{Origin and Structure of the Universe}, and
    CERN-RFBR grant 08-02-91009.}\hspace*{0.5em}
  on behalf of the COMPASS collaboration \\
  CERN, Geneva, Switzerland \\
  {\footnotesize On leave of absence from} \\
  Physik-Department E18, Technische Universit\"at M\"unchen, Garching, Germany \\
  E-mail: \email{bgrube@tum.de}}

\abstract{%
  The COmmon Muon and Proton Apparatus for Structure and Spectroscopy
  (COMPASS) is a multi-purpose fixed-target experiment at the CERN
  Super Proton Synchrotron (SPS) aimed at studying the structure and
  spectrum of hadrons.

  In the na\"ive Constituent Quark Model (CQM) mesons are bound states
  of quarks and antiquarks. QCD, however, predict the existence of
  hadrons beyond the CQM with exotic properties interpreted as excited
  glue (hybrids) or even pure gluonic bound states (glueballs). One
  main goal of COMPASS is to search for these states. Particularly
  interesting are so called spin-exotic mesons which have \jpc\
  quantum numbers forbidden for ordinary \Pqq\Pqqbar\ states.

  Its large acceptance, high resolution, and high-rate capability make
  the COMPASS experiment an excellent device to study the spectrum of
  light-quark mesons in diffractive and central production reactions
  up to masses of about 2.5\gevcc. COMPASS is able to measure final
  states with charged as well as neutral particles, so that resonances
  can be studied in different reactions and decay channels.

  During 2008 and 2009 COMPASS acquired large data samples using
  negative and positive secondary hadron beams on $\ell$H$_2$, Ni, and
  Pb targets. The presented overview of the first results from this
  data set focuses in particular on the search for spin-exotic mesons
  in diffractively produced \threepion, \Peta\Ppi, \Petaprime\Ppi, and
  \fivepion\ final states and the analysis of central-production of
  \twopion\ pairs in order to study glueball candidates in the scalar
  sector.
}

\FullConference{Xth Quark Confinement and the Hadron Spectrum \\
  8--12 October 2012 \\
  TUM Campus Garching, Munich, Germany}

\begin{document}

\section{Experimental Setup}

COMPASS is a two-stage high-resolution spectrometer that covers a wide
range of scattering angles and particle momenta~\cite{compass}. Both
stages of the spectrometer are equipped with hadronic and
electromagnetic calorimeters so that final states with charged as well
as neutral particles can be reconstructed. A Ring-Imaging Cherenkov
Detector (RICH) in the first stage can be used for particle
identification. It is able to separate kaons from pions up to momenta
of about 50\gevc. The target is surrounded by a Recoil Proton Detector
(RPD) that measures the time of flight of recoil protons using two
scintillator barrels. COMPASS uses the M2 beam line of the SPS which
can deliver secondary hadron beams with a momentum of up to 280\gevc\
and a maximum intensity of \tenpow[5]{7}~$\text{s}^{-1}$. The negative
hadron beam that was used for the analyses presented here has a
momentum of 190\gevc\ and consists of 97~\% \Ppim, 2~\% \PKm, and 1~\%
\Ppbar\ at the COMPASS target. Two ChErenkov Differential counters
with Achromatic Ring focus (CEDAR) upstream of the target are used to
identify the incoming beam particles.

\section{Search for Spin-Exotic Mesons in Diffractive Dissociation}

Diffractive dissociation reactions are known to exhibit a rich
spectrum of produced intermediate states. In the past several
candidates for spin-exotic mesons have been reported in pion-induced
diffraction~\cite{exotic_mesons}. In diffractive events the beam
hadron is excited to some intermediate state $X$ via $t$-channel
Reggeon exchange with the target. At 190\gevc\ beam momentum Pomeron
exchange is dominant. The $X$ decays into a $n$-body final state which
is detected by the spectrometer. The process $\text{beam} +
\text{target} \to X + \text{recoil}$, where $X \to h_1 \ldots h_n$, is
characterized by two kinematic variables: the square of the total
center-of-mass energy, $s$, and the squared four-momentum transfer to
the target, $t = (p_\text{beam} - p_X)^2$. It is customary to use the
variable $t' \equiv \abs{t} - \abs{t}_\text{min}$ instead of $t$,
where $\abs{t}_\text{min}$ is the minimum value of $\abs{t}$ for a
given invariant mass of $X$.

\subsection{$\pmb{\threepion}$ Final States from $\pmb{\Ppim}$ Diffraction}

A partial-wave analysis (PWA) of \threepion\ final states produced in
\Ppim\ diffraction on a Pb target in the squared four-momentum
transfer range of $0.1 < t' < 1.0$\gevcsq\ and extracted from data
taken during the pilot run in 2004 showed significant intensity in the
spin-exotic $\jpc = 1^{-+}$ partial wave in the $\Prz\Ppim$ decay
channel~\cite{compass_exotic}. A Breit-Wigner fit yielded resonance
parameters consistent with the disputed $\Ppione(1600)$ claimed in
this channel by other experiments~\cite{exotic}.

In 2008 COMPASS has acquired a large data set of the diffractive
dissociation reaction $\Ppim\,\Pp \to \threepion\,\Pp_\text{slow}$.
The trigger included a beam definition and the RPD, which ensured that
the target proton stayed intact and also introduced a lower bound for
$t'$ of about 0.1\gevcsq.
Diffractive events were enriched by an exclusivity cut around the
nominal beam energy. After all cuts the analyzed \threepion\ sample
contains about \tenpow[5]{7}~events in the range $0.1 < t' <
1.0$\gevcsq.

The PWA approach employs the isobar model~\cite{isobar} in which the
decay $X^- \to \threepion$ is decomposed into a chain of successive
two-body decays. The $X^-$ with quantum numbers \jpc\ and spin
projection $M^\epsilon$ is assumed to decay into a \twopion\
resonance, the so-called isobar, and a bachelor pion. The isobar has
spin $S$ and a relative orbital angular momentum $L$ with respect to
$\Ppim_\text{bachelor}$. A partial wave is thus defined by $\jpc
M^\epsilon[\text{isobar}]L$, where $\epsilon = \pm 1$ is the
reflectivity which corresponds to the naturality of the exchanged
particle in the production process~\cite{reflectivity}.

The spin-density matrix for the chosen set of 52 partial waves plus an
incoherent isotropic background wave is determined by unbinned
extended maximum likelihood fits performed in 20\mevcc\ wide bins of
the three-pion invariant mass $m_X$. In these fits no assumption is
made on the produced resonances $X^-$ other than that their production
strength is constant within a $m_X$ bin. The PWA model includes five
\twopion\ isobars~\cite{hadron2011_florian}:
$(\Ppi\Ppi)_\text{$S$-wave}$, $\Pr(770)$, $\Pfzero(980)$,
$\Pftwo(1270)$, and $\Prthree(1690)$. They were described using
relativistic Breit-Wigner line shape functions including
Blatt-Weisskopf barrier penetration factors~\cite{bwFactor}. For the
\twopion\ $S$-wave we use the parametrization from~\cite{vesSigma}
with the $\Pfzero(980)$ subtracted from the elastic $\Ppi\Ppi$
amplitude and added as a separate Breit-Wigner resonance. Mostly
natural-parity waves are needed to describe the data. A rank-two
spin-density matrix was used in order to account for spin-flip and
spin-non-flip amplitudes at the target vertex.

\begin{figure}[t]
  \vspace*{-3ex}
  \includegraphics[width=0.31\textwidth]{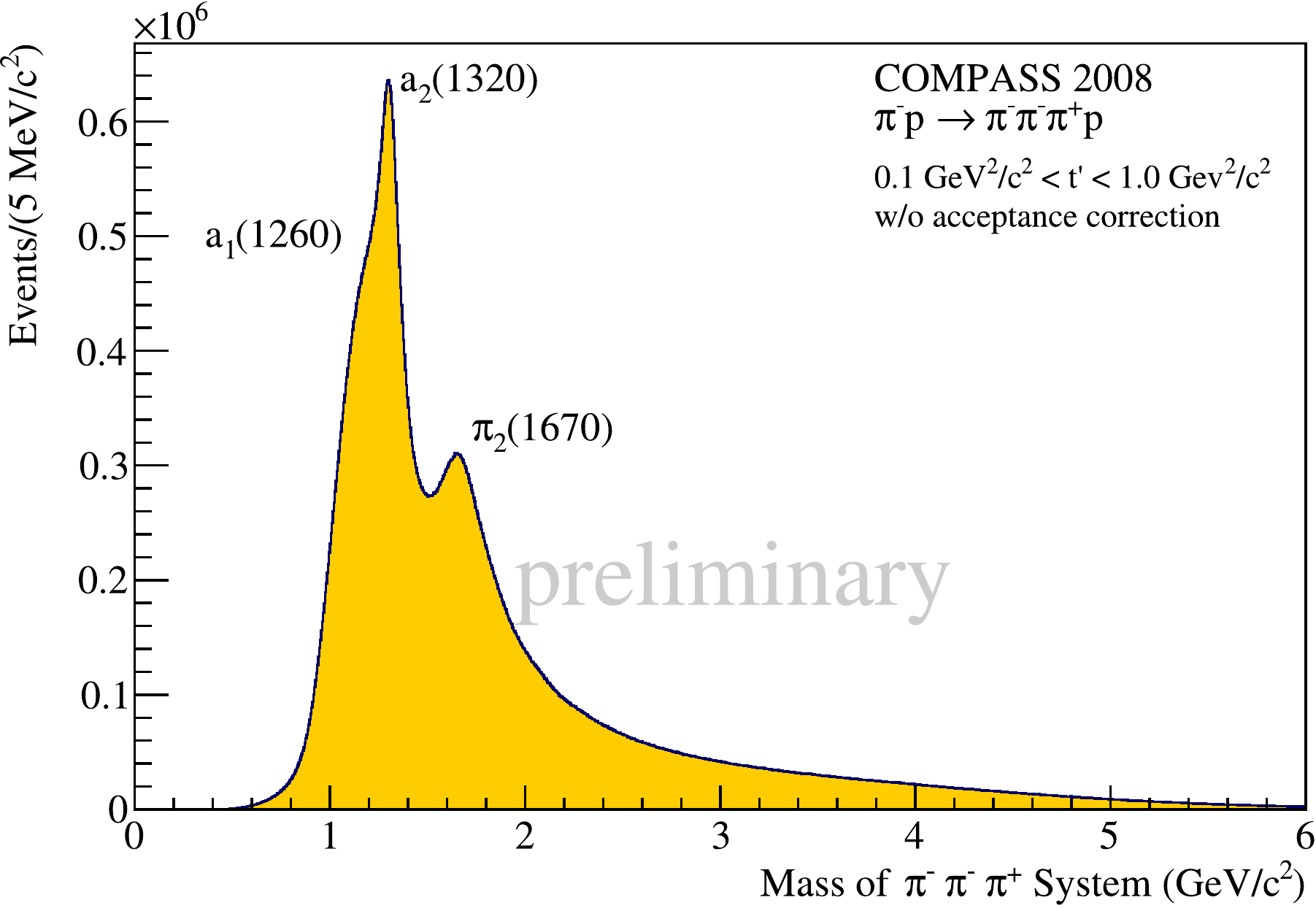} \quad
  \includegraphics[width=0.31\textwidth]{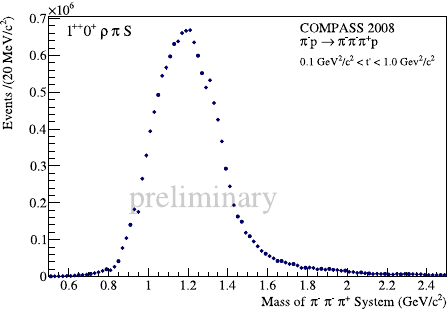} \quad
  \includegraphics[width=0.31\textwidth]{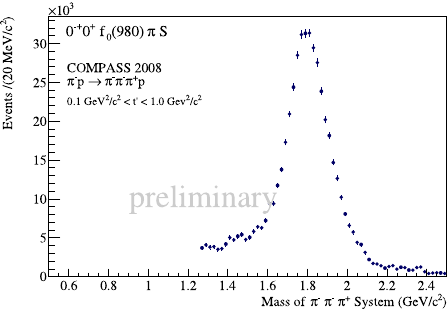} \\
  \includegraphics[width=0.31\textwidth]{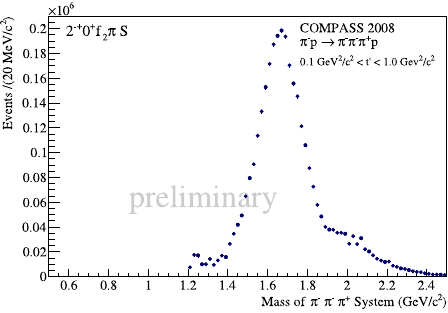} \quad
  \includegraphics[width=0.31\textwidth]{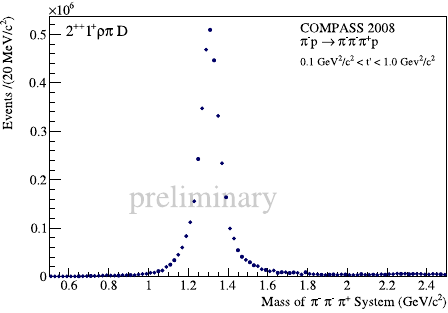} \quad
  \includegraphics[width=0.31\textwidth]{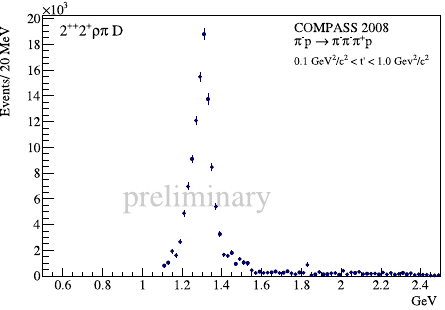}
  \vspace*{-2ex}
  \caption{\emph{Left and center columns:} \threepion\ invariant mass
    spectrum (top left). Intensities of the three major waves:
    \protect \wavespec{1}{+}{+}{0}{+}{\Pr\Ppi}{S} wave with
    the $\Paone(1260)$ (top center), \protect
    \wavespec{2}{+}{+}{1}{+}{\Pr\Ppi}{D} wave with the $\Patwo(1320)$
    (bottom center), and \protect
    \wavespec{2}{-}{+}{0}{+}{\Pftwo\Ppi}{S} wave with the $\Ppitwo(1670)$
    (bottom left). \emph{Right column:} Examples for small waves (note
    the different $y$ scale): \protect
    \wavespec{0}{-}{+}{0}{+}{\Pfzero(980)\Ppi}{S} wave with the $\Ppi(1800)$ (top right) and
    \protect \wavespec{2}{+}{+}{2}{+}{\Pr\Ppi}{D} with the $\Patwo(1320)$ (bottom right).}
  \label{fig:2008_3pi}
\end{figure}

The intensities of the three dominant waves in the \threepion\ final
state, \wavespec{1}{+}{+}{0}{+}{\Pr\Ppi}{S},
\wavespec{2}{+}{+}{1}{+}{\Pr\Ppi}{D}, and
\wavespec{2}{-}{+}{0}{+}{\Pftwo\Ppi}{S}, are shown in
\figref{2008_3pi} left and center. They contain resonant structures
that correspond to the well-known $\Paone(1260)$, $\Patwo(1320)$, and
$\Ppitwo(1670)$, respectively~\cite{hadron2011_florian}. That the
applied analysis technique is able to cleanly separate even small
waves with intensities at the percent level of that of the dominant
waves is illustrated in \figref{2008_3pi} right. It shows two
exemplary waves, \wavespec{2}{+}{+}{2}{+}{\Pr\Ppi}{D} and
\wavespec{0}{-}{+}{0}{+}{\Pfzero(980)\Ppi}{S}, which exhibit clear
peaks of the $\Patwo(1320)$ (this time with $M = 2$) and the
$\Ppi(1800)$, respectively.

\begin{figure}[t]
  \vspace*{-3ex}
  \includegraphics[width=0.31\textwidth]{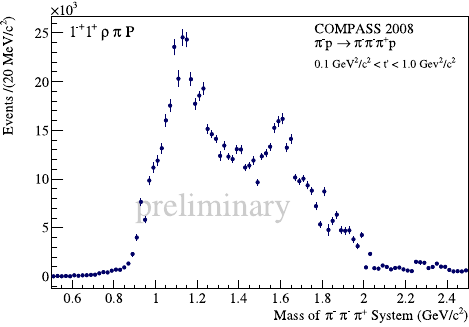} \quad
  \includegraphics[width=0.31\textwidth]{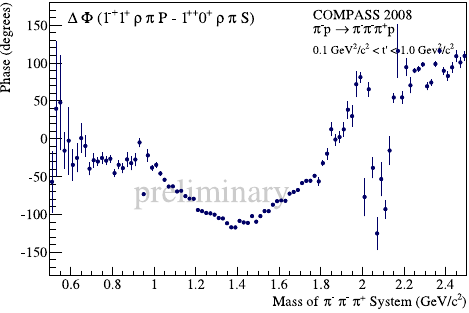} \quad
  \includegraphics[width=0.31\textwidth]{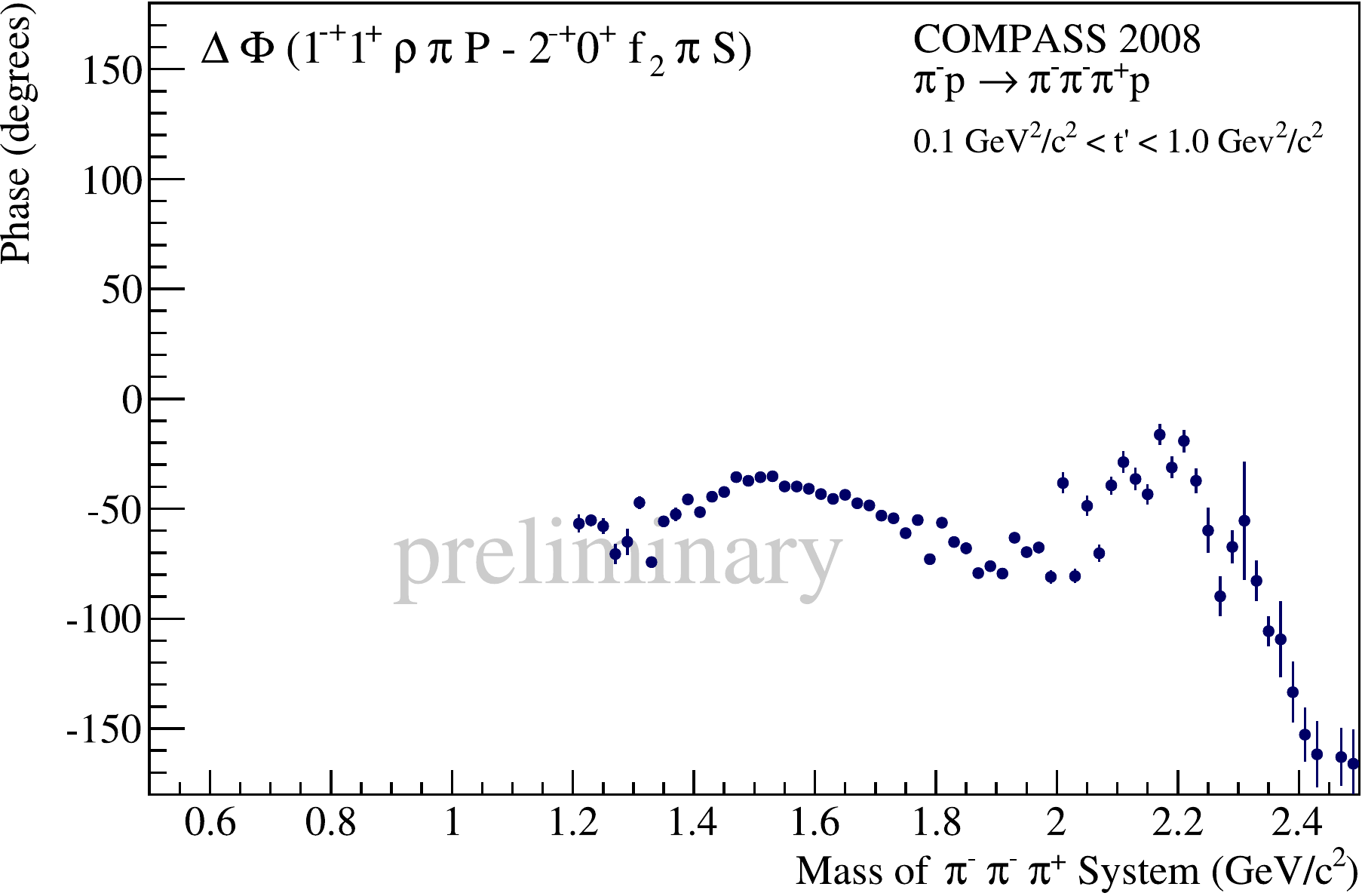}
  \vspace*{-2ex}
  \caption{\emph{Left:} Intensity of the spin-exotic \protect
    \wavespec{1}{-}{+}{1}{+}{\Pr\Ppi}{D} wave. \emph{Center:} Phase of
    the $1^{-+}$ wave relative to the \protect
    \wavespec{1}{+}{+}{0}{+}{\Pr\Ppi}{S} wave. \emph{Right:} Relative phase
    with respect to the \protect
    \wavespec{2}{-}{+}{0}{+}{\Pftwo\Ppi}{S} wave.}
  \label{fig:2008_3pi_exotic}
\end{figure}

\Figref{2008_3pi_exotic} left shows the intensity of the spin-exotic
\wavespec{1}{-}{+}{1}{+}{\Pr\Ppi}{P} wave. The bump around 1.2\gevcc\
does not seem to be of resonant nature. It is unstable with respect to
changes in the PWA model which hints that it is rather an artifact of
the analysis method.
On the other hand the peak structure around 1.6\gevcc\ as well as the
corresponding rising phase motion with respect to the tail of the
$\Paone(1260)$ in the \wavespec{1}{+}{+}{0}{+}{\Pr\Ppi}{S} wave (cf.
\figref{2008_3pi_exotic} center) are stable against modifications of
the fit model. As \figref{2008_3pi_exotic} right shows, the structure
is phase locked with the $\Ppitwo(1670)$ in the
\wavespec{2}{-}{+}{0}{+}{\Pftwo\Ppi}{S} wave. This is consistent with
the results obtained from a PWA of the pilot-run data taken with a Pb
target~\cite{compass_exotic}. The interpretation of the $1^{-+}$ wave
in terms of resonances, however, is still unclear. There seem to be
significant contributions from non-resonant Deck-like processes that
need to be included into the fit model.

The partial waves also exhibit different $t'$ behavior depending on
their spin projection $M$ and their resonance contents. This is
problematic, since the data cover a rather wide $t'$ range from 0.1 to
1.0\gevcsq\ while at the same time the PWA model assumes full
coherence of all partial waves. This issue will be addressed by
performing a two-dimensional PWA in bins of $m_X$ and $t'$.

\subsection{$\pmb{\Ppim\Peta}$ and $\pmb{\Ppim\Petaprime}$  Final States from $\pmb{\Ppim}$ Diffraction}

Previous experiments claimed spin-exotic $\jpc = 1^{-+}$ resonances
also in the $\Ppim\Peta$~\cite{eta_etaprime_pi, eta_pi}
and $\Ppim\Petaprime$~\cite{eta_etaprime_pi, etaprime_pi} final
states. However, the resonant nature of the observed signals is still
controversial~\cite{adam}. COMPASS performed a PWA of both final
states in the diffractive reaction $\Ppim\,\Pp \to
\Ppim\Peta\raisebox{1ex}{$\scriptstyle($}\vphantom{\Peta}'\raisebox{1ex}{$\scriptstyle)$}\,\Pp_\text{slow}
\to \threepion\,\PPh\PPh\,\Pp_\text{slow}$ in the kinematic range $0.1
< t' < 1.0\gevcsq$~\cite{qnp_tobias}. The \Peta\ is reconstructed via
its decay to \Ppip\Ppim\Ppiz\ where the \Ppiz\ goes to a photon
pair. In order to reconstruct the \Petaprime, \Peta\ reconstructed in
the \PPh\PPh\ channel are combined with a $\Ppip\Ppim$ pair. The final
selected data sample contains about 100\,000 \Ppim\Peta\ and 35\,000
\Ppim\Petaprime\ events.

The performed PWA follows previous analyses and includes $S$, $P$, and
$D$ waves with $M \leq 1$ and both natural and unnatural parity
exchange. In addition a $4^{++}\, 1^+$ $G$ wave and an incoherent
isotropic background wave were included. For both channels the bulk of
the intensity is described by the $1^{-+}$, $2^{++}$, and $4^{++}$
waves with $M = 1$ and natural parity exchange.

\begin{figure}[t]
  \vspace*{-3ex}
  \includegraphics[width=0.31\textwidth]{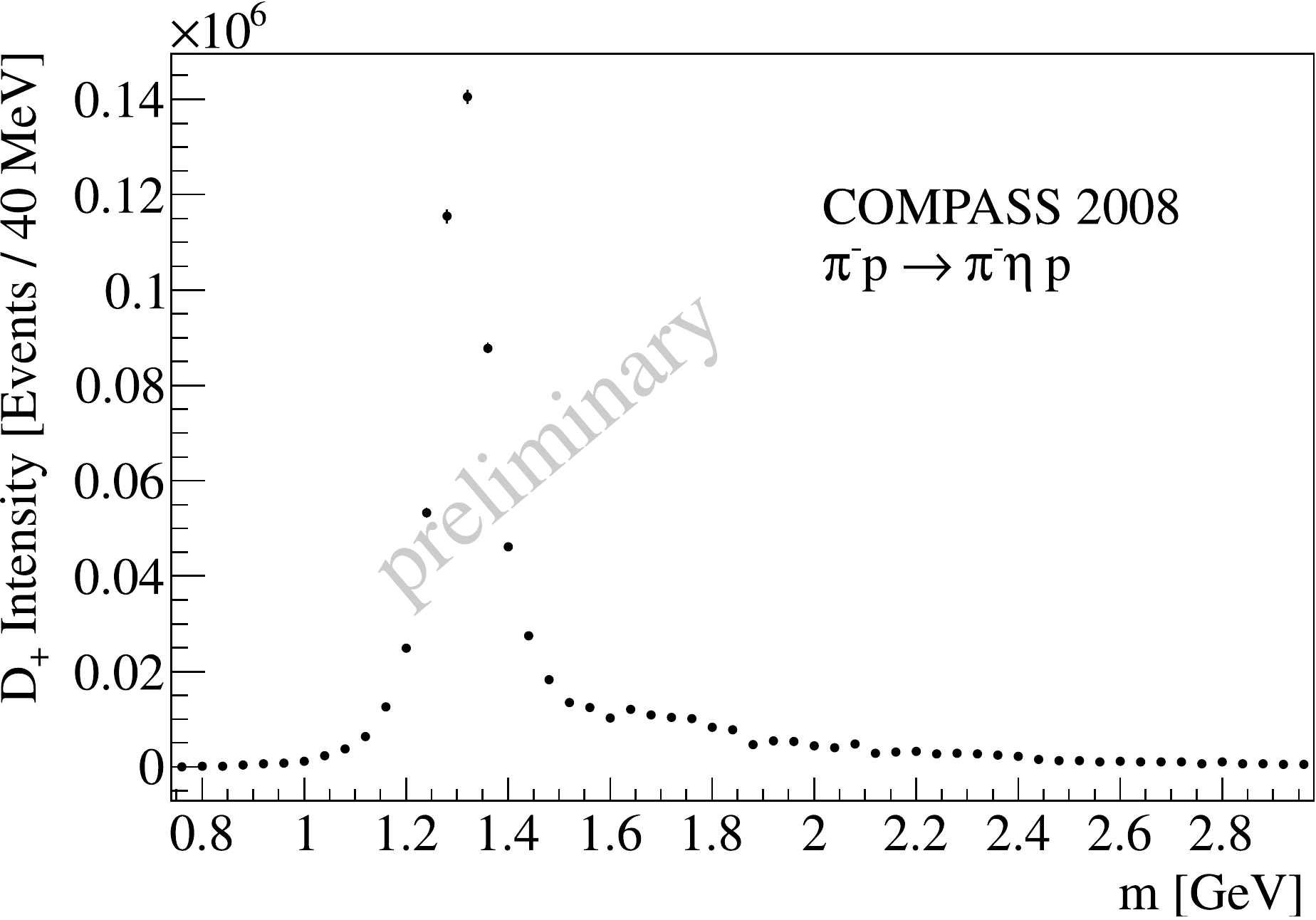} \quad
  \includegraphics[width=0.31\textwidth]{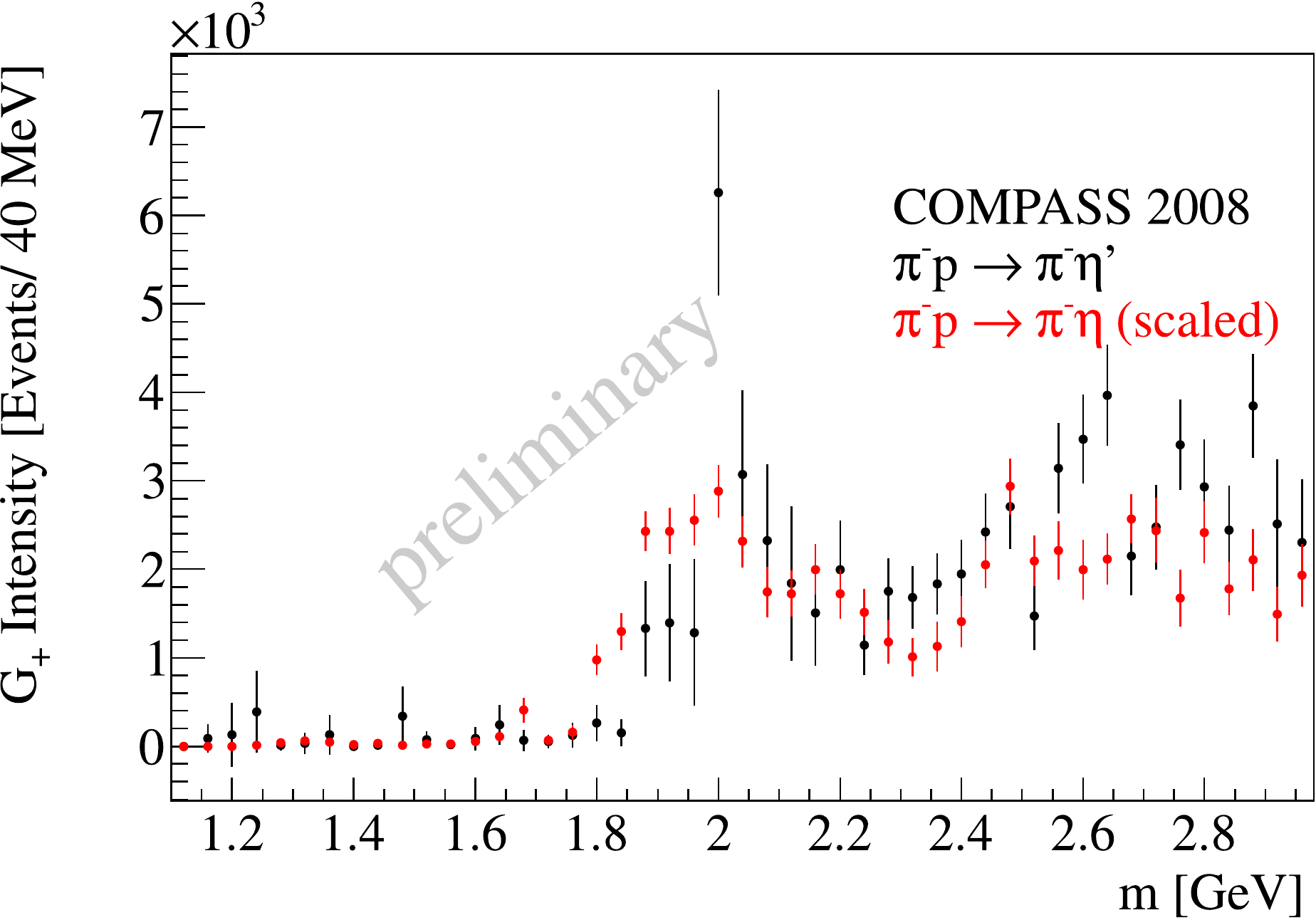} \quad
  \includegraphics[width=0.31\textwidth]{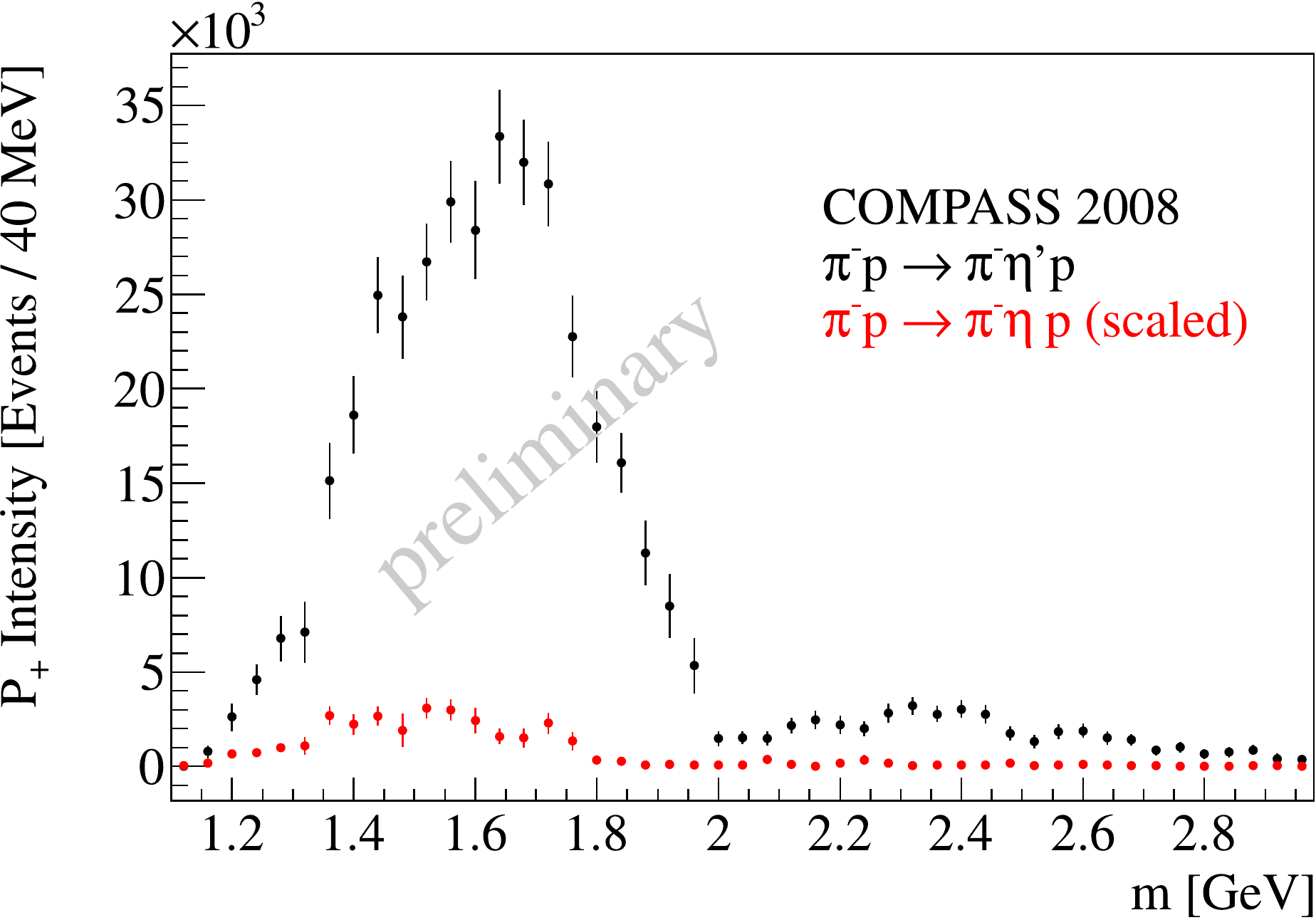} \\
  \includegraphics[width=0.31\textwidth]{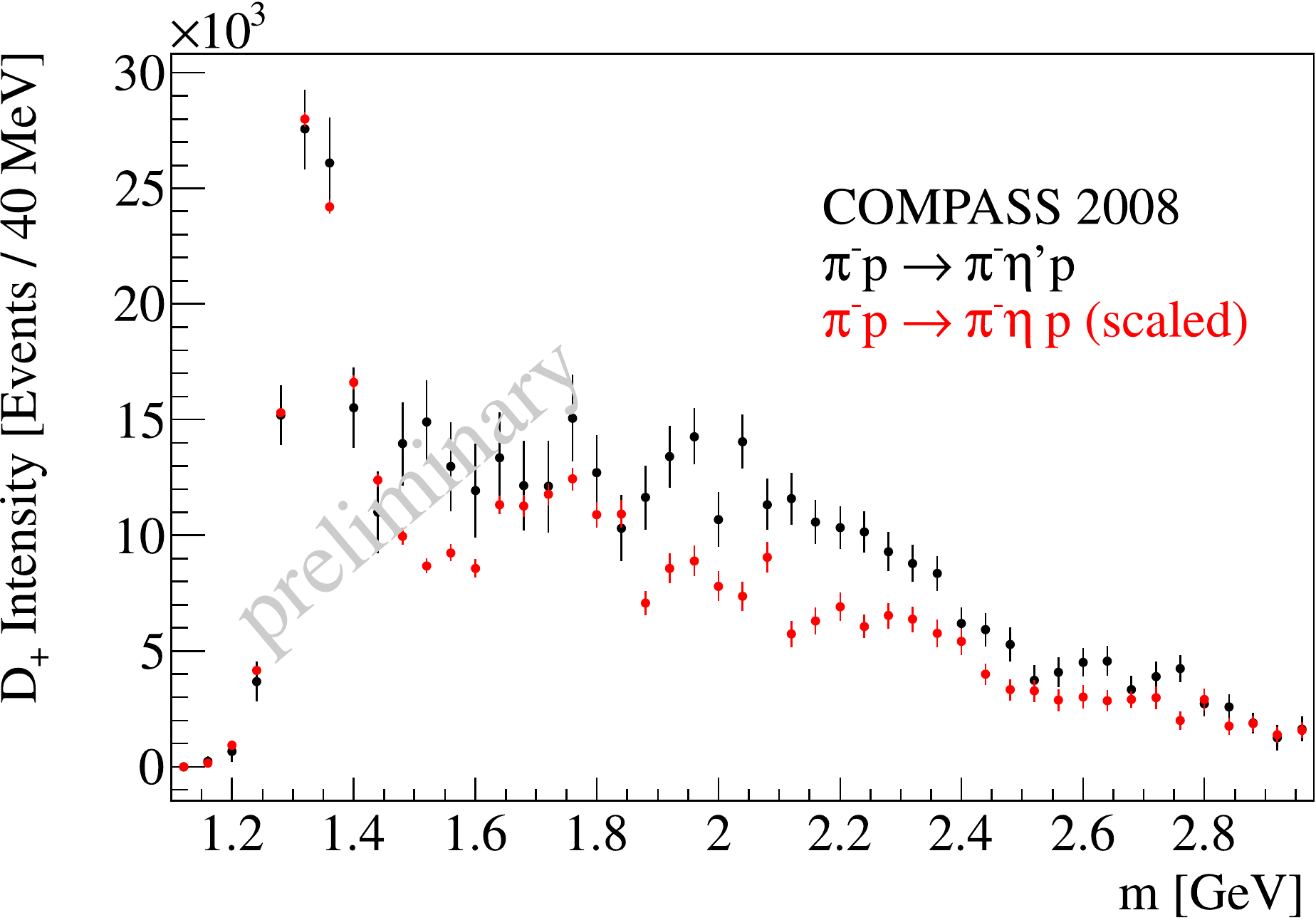} \quad
  \includegraphics[width=0.31\textwidth]{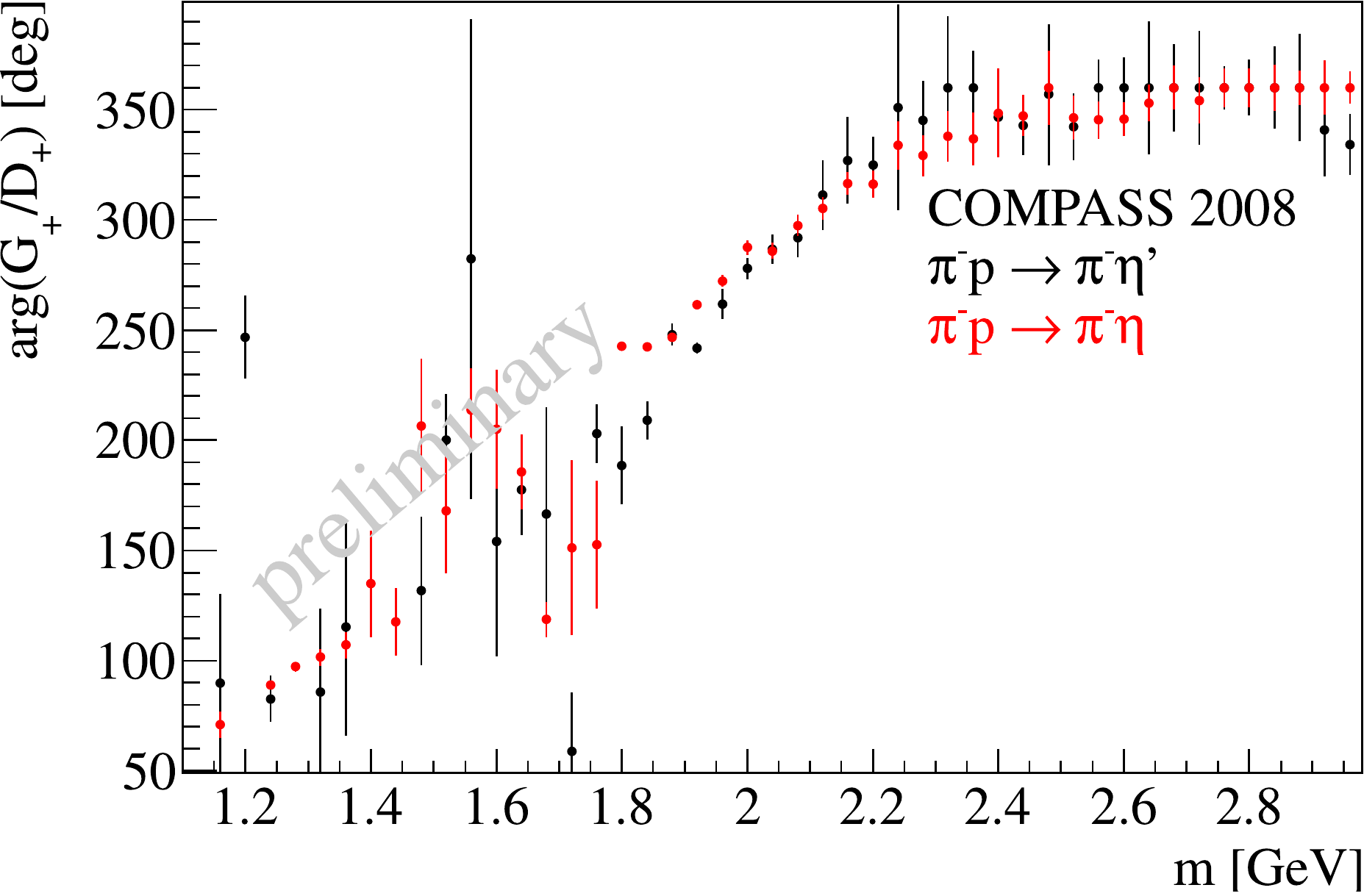} \quad
  \includegraphics[width=0.31\textwidth]{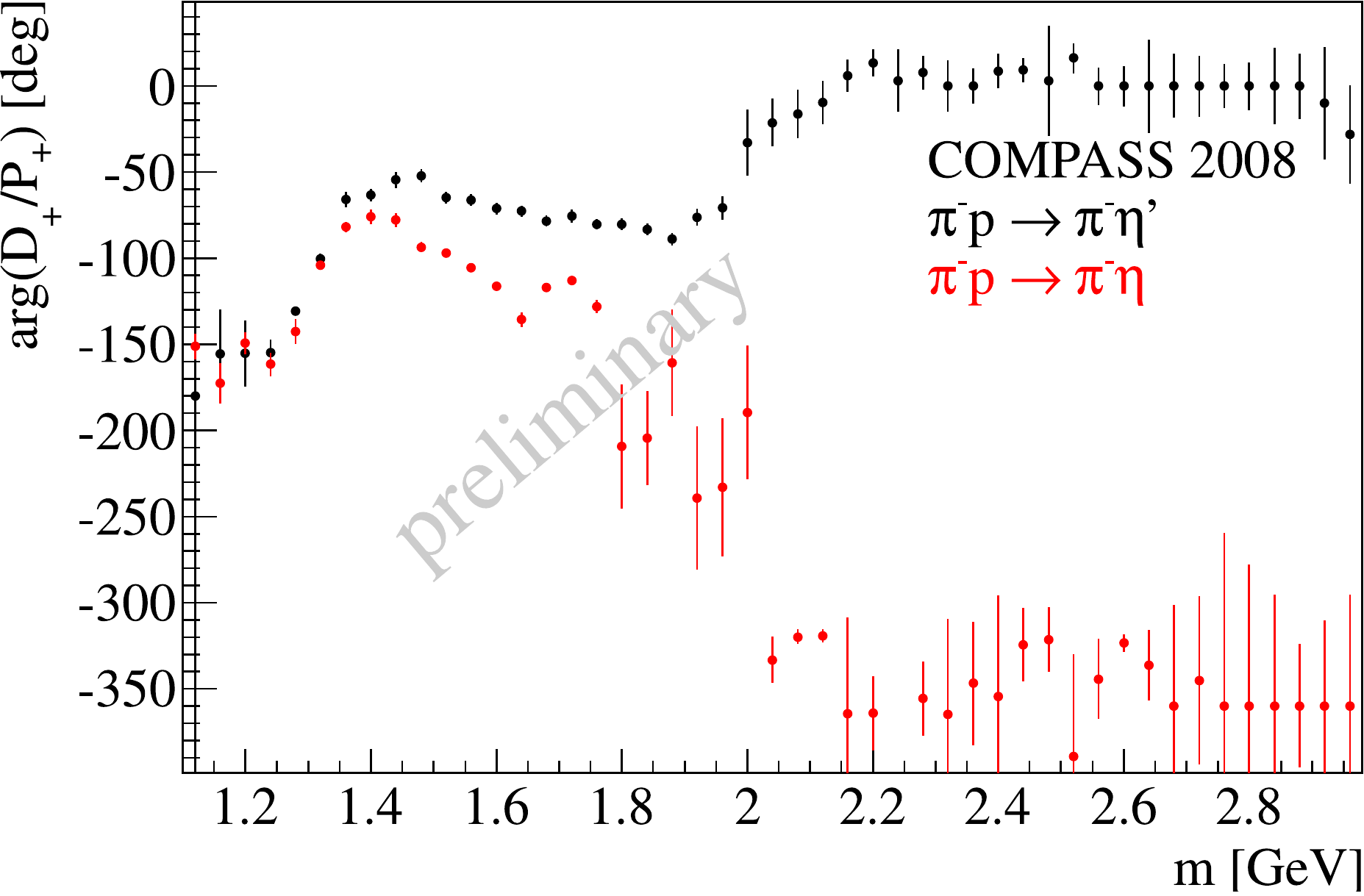}
  \vspace*{-2ex}
  \caption{\emph{Left column:} Intensity of the $2^{++}\, 1^+$ ($D_+$)
    wave in the \Ppim\Peta\ (top) and \Ppim\Petaprime\ (bottom, black)
    channel, respectively. The red dots in the bottom plot show the
    \Ppim\Peta\ $D_+$ intensity scaled by phase space.  \emph{Center
      and right column:} Intensities and relative phases for
    \Ppim\Petaprime\ (black) and \Ppim\Peta\ scaled by phase space
    (red): $4^{++}\, 1^+$ ($G_+$) intensity (top center), spin-exotic
    $1^{-+}\, 1^+$ ($P_+$) intensity (top right), relative phases $G_+ -
    D_+$ (bottom center) and $D_+ - P_+$ (bottom right).}
  \label{fig:2008_eta_etaprime_pwa}
\end{figure}

The left column of \figref{2008_eta_etaprime_pwa} shows the $2^{++}\,
1^+$ ($D_+$) intensity for the two final states. Although the two
distributions look rather different they are related to each other by
the fact that \Peta\ and \Petaprime\ are mixtures of the basis states
$\Peta_\Pqn = (1 / \sqrt{2})\, (\Pqu\Pqubar + \Pqd\Pqdbar)$ and
$\Peta_\Pqs = \Pqs\Pqsbar$. The partial-wave amplitudes $T_J$ of a
spin-$J$ resonance with negligible \Pqs\Pqsbar\ content decaying into
\Ppim\Peta\ and \Ppim\Petaprime\ should be related to each other by
the \Peta-\Petaprime\ pseudoscalar mixing angle $\phi$ in the flavor
basis, phase space, and barrier penetration
factors~\cite{eta-etaprime_mixing}. With $q$ being the two-body
breakup momentum the amplitude ratio can be written as
\begin{equation}
  \frac{T_J^{\Ppi\Peta'}(m_X)}{T_J^{\Ppi\Peta}(m_X)}
  = \tan \phi\, \lrBrk{\frac{q^{\Ppi\Peta'}(m_X)}{q^{\Ppi\Peta}(m_X)}}^{J + 1/2}
  \label{eq:eta_etaprime_ratio}
\end{equation}
Here simplified barrier factors of the form $q^J$ were used.

The bottom left plot in \figref{2008_eta_etaprime_pwa} shows in red
the $D_+$ intensity of the \Ppim\Peta\ channel scaled by the phase
space factor from \equref{eta_etaprime_ratio} overlaid on top of the
corresponding intensity in the \Ppim\Petaprime\ final state. The phase
space scaling leads to a remarkable similarity of the \Ppim\Peta\ and
\Ppim\Petaprime\ intensities. This behavior is expected for
light-quark resonances like the $\Patwo(1320)$, but it is rather
astonishing that the scaling also holds in the high-mass region where
non-resonant contributions are expected to dominate. As the middle row
of \figref{2008_eta_etaprime_pwa} shows, the same behavior is observed
for the $4^{++}\, 1^+$ ($G_+$) partial waves with the
$\Pafour(2040)$. Also the relative phase between the $G_+$ and the
$D_+$ waves is very similar in both channels which means that the
physical composition of the two waves is almost identical.

The picture is much different for the spin-exotic $1^{-+}\, 1^+$
($P_+$) wave which is the dominant wave in the \Ppim\Petaprime\
channel with a prominent broad structure around 1.6\gevcc\ whereas it
is strongly suppressed in the \Ppim\Peta\ channel
(cf. \figref{2008_eta_etaprime_pwa} right). This observation is
consistent with previous experiments and with the suspected
non-\Pqq\Pqqbar\ character of this wave~\cite{eta-etaprime_ratio}. The
$P_+$ wave exhibits a slow phase motion with respect to the $D_+$ wave
in the 1.6\gevcc\ mass region which, however, evolves differently in
the two final states. This is a hint that the resonant content is
different.
The resonance interpretation of the $P_+$ wave requires a better
understanding of the high-mass region of the $D_+$ and $G_+$ waves
against which the phase motion of the $P_+$ is measured. In this
region non-resonant contributions from double-Regge processes are
expected to play an important role which should be included in the fit
model~\cite{confx_vincent}.

\subsection{$\pmb{\fivepion}$  Final States from $\pmb{\Ppim}$ Diffraction}

The diffractive reaction $\Ppim \text{Pb} \to \fivepion\, \text{Pb}$
is very interesting, because the 5\Ppi\ final state gives access to
the mass region around and above 2\gevcc\ which is also called the
``light-meson frontier''. Only little is known about resonances in
this region. There are many missing states as well as states that need
confirmation.

\begin{table}[t]
  \vspace*{-2ex}
  \begin{center}
  \begin{small}
    \begin{tabular}{lll}
      $\pmb{\twopion}$ \textbf{Isobars} &
      $\pmb{\Ppim\Ppip\Ppipm}$ \textbf{Isobars} &
      $\pmb{\fourpion}$ \textbf{Isobars} \\
      \hline
      $(\Ppi\Ppi)_{S\text{-wave}}$, $\Pr(770)$ &
      $\Paone(1260)$, $\Patwo(1320)$ &
      $\Pftwo(1270)$, $\Pfone(1285)$, $\Pfzero(1370, 1500)$, $\Pr'(1450, 1700)$ \\
    \end{tabular}
  \end{small}
  \end{center}
  \vspace*{-3ex}
  \caption{Isobars used in the wave set of the \fivepion\ PWA~\cite{thesis_sebastian}.}
  \label{tab:5pi_isobars}
\end{table}

A PWA of 200\,000 exclusive events was performed in the kinematic
region $t' < \tenpow[5]{-3}\gevcsq$. Since a Pb target was used the
recoil could not be detected. The partial-wave decomposition of
diffractively produced \fivepion\ final states is challenging.  Due to
the lack of dominant structures in the 5\Ppi\ invariant mass
distribution and that of its subsystems, it is difficult to constraint
the wave set which is potentially large, because of the many allowed
isobars. In order
\begin{wraptable}{r}{0.34\textwidth}
  \vspace*{-3ex}
  \begin{center}
  \begin{small}
    \begin{tabular}{llr}
      \textbf{Resonance} & & \textbf{Fit result} \\
                         & & $\pmb{\text{MeV}\! / c^2}$ \\
      \hline
      $\Ppi(1300)$    & $M$         & $1400$ (at limit) \\
                      & $\varGamma$ & $500$ (fixed) \\
      $\Ppi(1800)$    & $M$         & ${1781 \pm 5}^{+8}_{-6}$ \\
                      & $\varGamma$ & ${168 \pm 9}^{+62}_{-15}$ \\
      $\Paone(1900)$  & $M$         & ${1853 \pm 7}^{+36}_{-49}$ \\
                      & $\varGamma$ & ${443 \pm 14}^{+98}_{-65}$ \\
      $\Paone(2200)$  & $M$         & ${2202 \pm 8}^{+53}_{-11}$ \\
                      & $\varGamma$ & ${402 \pm 17}^{+125}_{-51}$ \\
      $\Ppitwo(1670)$ & $M$         & $1719.0$ (fixed) \\
                      & $\varGamma$ & $251.4$ (fixed) \\
      $\Ppitwo(1880)$ & $M$         & ${1854 \pm 6}^{+6}_{-9}$ \\
                      & $\varGamma$ & ${259 \pm 13}^{+7}_{-31}$ \\
      $\Ppitwo(2100)$ & $M$         & ${2133 \pm 12}^{+43}_{-18}$ \\
                      & $\varGamma$ & ${448 \pm 22}^{+80}_{-40}$ \\
    \end{tabular}
  \end{small}
  \end{center}
  \vspace*{-3ex}
  \caption{Summary of extracted $5\Ppi$ resonance
    parameters~\cite{thesis_sebastian}. The first uncertainty is
    statistical, the second represents the systematic error including
    uncertainties from the choice of the final wave set.}
  \label{tab:5pi_fit}
\end{wraptable}
to explore the model
space more systematically an evolutionary algorithm was developed that
uses a Bayesian goodness-of-fit criterion that takes into account the
model complexity to iteratively find the best set of
waves~\cite{thesis_sebastian}. A pool of 284 waves was offered to the
algorithm. The best model found consists of 31 partial wave plus an
incoherent isotropic background wave. \Tabref{5pi_isobars} lists the
used isobars. An advantage of this automatized method is that it not
only finds a wave set but also gives an estimate of the systematic
uncertainty introduced by the particular choice of the wave set.

\begin{figure}[t]
  \vspace*{-3ex}
  \begin{center}
    \includegraphics[width=\textwidth]{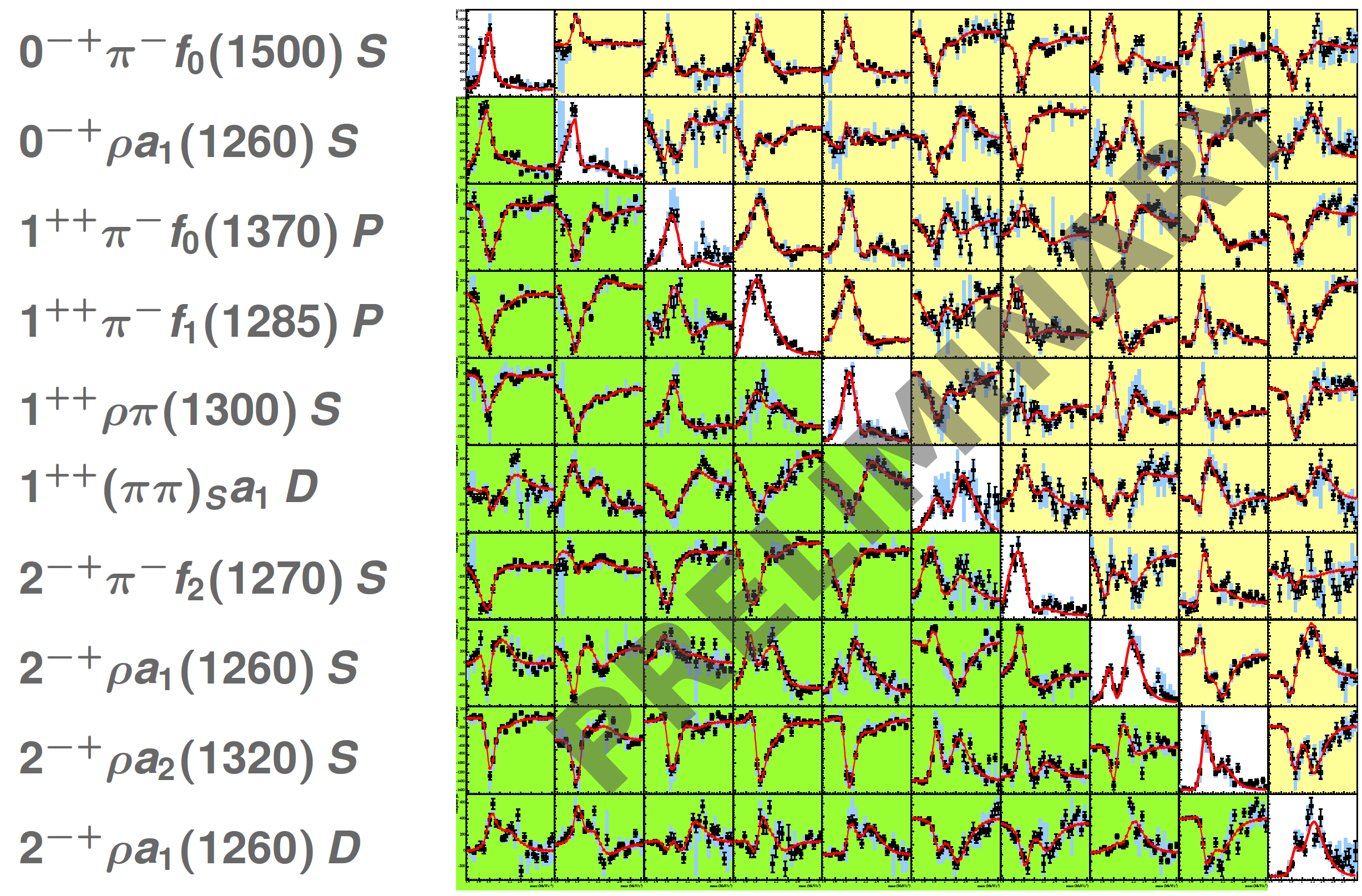}
  \end{center}
  \vspace*{-4ex}
  \caption{Fit of the $m_{5\Ppi}$ dependence of a spin-density
    submatrix of 10 waves~\cite{thesis_sebastian}. The diagonal
    elements show the wave intensities, the off-diagonal ones the
    interference terms (top right triangle: real, bottom left
    triangle: imaginary part). The light blue error bars represent the
    uncertainty from the choice of the wave set.}
  \label{fig:5pi_fit}
\end{figure}

The mass dependence of a submatrix of the spin-density matrix
consisting of 10 out of the 31 waves was parametrized using a simple
model that consists of 6 resonances that were described by
constant-width relativistic Breit-Wigners plus background
terms. Although mixing and coupled-channel effects were neglected the
model is able describe the data surprisingly well as is illustrated in
\figref{5pi_fit}. \Tabref{5pi_fit} summarizes the extracted resonance
parameters. Being the first analysis of diffractively produced
\fivepion, the results of this PWA should be interpreted with
care. The analysis is still based on a number of assumptions. In
particular the lack of knowledge on the properties of the four-pion
isobars is difficult to quantify.

\section{Search for Scalar Glueball Candidates in Central Production}

\begin{wrapfigure}{r}{0.35\textwidth}
  \vspace*{-2ex}
  \includegraphics[width=\linewidth]{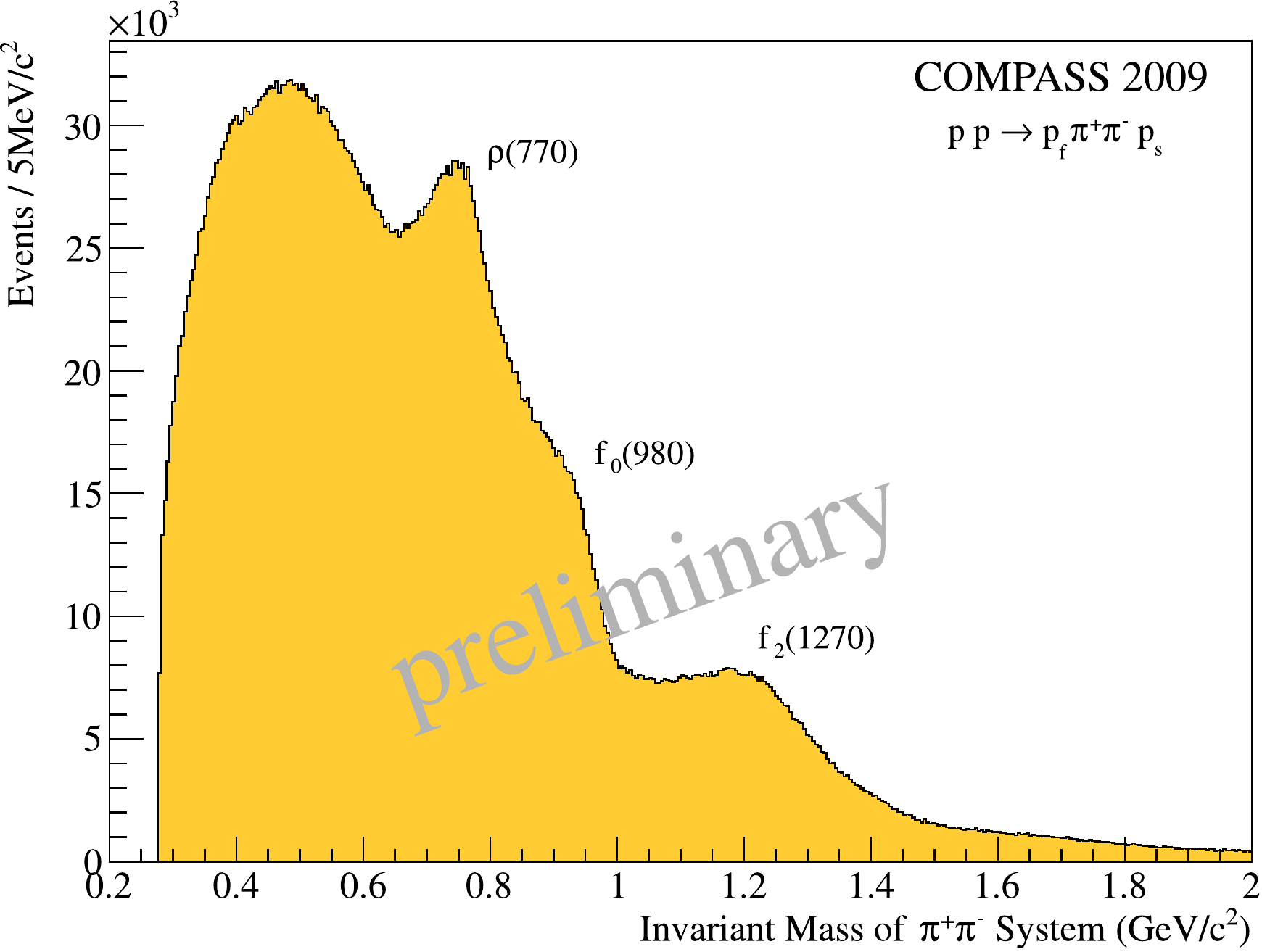}
  \vspace*{-5ex}
  \caption{\twopion\ invariant mass distribution.}
  \label{fig:cp_twopion_mass}
\end{wrapfigure}
Pomeron-Pomeron fusion processes are believed to provide a glue-rich
environment which should lead to an enhanced cross section for the
production of glueballs. Although glueball candidates are discussed in
the literature, their existence could not yet be confirmed
experimentally. Continuing the efforts of CERN OMEGA spectrometer in
the late 1990s~\cite{omega} COMPASS studies the central-production
reaction $\Pp\Pp \to \Pp_\text{fast}\, \twopion\, \Pp_\text{slow}$
using a 190\gevc\ positive secondary hadron beam which consists of
75~\% \Pp, 24~\% \Ppip, and 1~\% \PKp\ at the $\ell$H$_2$
target~\cite{qnp_alex}.

Centrally produced \twopion\ are separated from beam-diffraction
events by cutting on the invariant mass $m(p_\text{fast}\Ppipm) >
1.5\gevcc$. Elastic scattering at the target vertex is ensured by the
RPD trigger. This selects \twopion\ pairs within $\abs{x_F} \leq
0.25$. At COMPASS energies in addition to Pomeron-Pomeron fusion also
Reggeon-Pomeron and Reggeon-Reggeon processes are expected to
contribute. This can be seen in the \twopion\ invariant mass spectrum
shown in \figref{cp_twopion_mass}, where in addition to structures
from $\Pfzero(980)$ and $\Pftwo(1270)$ a clear $\Pr(770)$ peak is
visible.

The employed PWA method is similar to the one used by
WA102~\cite{wa102_pipi}. The \twopion\ pair is assumed to be produced
in the collision of two exchanged objects emitted by the scattered
protons. The one emitted by the beam proton carries a squared
four-momentum transfer $t_1$, the one from the target proton $t_2$,
respectively. We make the strong assumption that exchange particle
$t_1$ only transmits helicity $\lambda = 0$ and that we can treat it
like an external particle.

A complication arises from the fact that the two-pseudoscalar final
state suffers from mathematical ambiguities, which means that for a
given wave set different partial-wave decompositions result in exactly
the same angular distribution. The PWA was performed using a wave set
consisting of $S$, $P$, and $D$ waves which leads to eight ambiguous
solutions. Additional constraints are needed to select the physical
solution. For six solutions, most of the intensity accumulates in a
single wave which is clearly unphysical, since at least one resonance
is known to be present in $S$, $P$ and $D$ waves
(cf. \figref{cp_twopion_mass}). The choice among the two remaining
solutions is not evident, the intensities and phases are very similar.
The remaining solutions are compatible with the physical
constraints. A clear $\Pftwo(1270)$ peak can be seen in the $2^{++}\,
0^-$ ($D_0^-$) intensity. It is accompanied by a phase motion with
respect to the $0^{++}\, 0^-$ ($S_0^-$) wave. The $P$ waves show
$\Pr(770)$ peaks. The most interesting wave is, of course, the
$S_0^-$. Its interpretation is, however, challenging and work in
progress.


\section{Summary}

COMPASS is a unique apparatus to study light-quark hadron spectroscopy
in diffractive and central production reactions. Large data sets were
collected for various final states with charged as well as neutral
particles using different target materials. The main focus of the
first analyses lies on the search for spin-exotic mesons. $\jpc =
1^{-+}$ signals are observed in diffractively produced 3\Ppi\ and
$\Ppi\Peta\raisebox{1ex}{$\scriptstyle($}\vphantom{\Peta}'\raisebox{1ex}{$\scriptstyle)$}$
final states. However, their interpretation in terms of resonances is
complicated by strong non-resonant contributions that need to be taken
into account in the analysis. In the future, the search for
spin-exotic signals will be extended to channels like $\Ppim\Po\Ppiz$
and $\Ppim(\PK\PKbar)^\pm\Ppimp$. Glueball candidates are studied in
centrally produced \twopion\ pairs. First PWA show promising results
and this analysis is beeing extended to $\PKp\PKm$, $\Ppiz\Ppiz$, and
$\PKs\PKsbar$ final states.


\begin{thebibliography}{99}

  \setlength{\parskip}{0ex}%
  \setlength{\itemsep}{0ex}%

\bibitem{compass} P. Abbon \others,
  Nucl. Instrum. Methods~\textbf{A577} (2007) 455; M. Alekseev
  \others, ``The COMPASS 2008 Spectrometer'' to be submitted to
  Nucl. Instrum. Methods~\textbf{A} (2013).


\bibitem{exotic_mesons} C. Meyer \others,
  Phys. Rev.~\textbf{C82} (2010) 025208;
  E. Klempt \others,
  Phys. Rept.~\textbf{454} (2007) 1.

\bibitem{compass_exotic} A. Alekseev \others,
  Phys. Rev. Lett.~\textbf{104} (2010) 241803.

\bibitem{exotic} G. S. Adams \others,
  Phys. Rev. Lett.~\textbf{81} (1998) 5760;
  Y. Khokhlov,
  Nucl. Phys.~\textbf{A663} (2000) 596;
  S. U. Chung \others,
  Phys. Rev.~\textbf{D65} (2002) 072001.

\bibitem{isobar} J. D. Hansen \others,
  Nucl. Phys.~\textbf{B81} (1974) 403.

\bibitem{reflectivity} S. U. Chung and T. L. Trueman,
  Phys. Rev.~\textbf{D11} (1975) 633.

\bibitem{hadron2011_florian} F. Haas, \textit{Proceedings of the XIV
    International Conference on Hadron Spectroscopy (hadron2011)},
  Munich (2011), eConf C110613 (2011) [arXiv:1109.1789].

\bibitem{bwFactor} F. von Hippel and C. Quigg,
  Phys. Rev.~\textbf{D5} (1972) 624.

\bibitem{vesSigma} D.V. Amelin \others,
  Phys. Lett.~\textbf{B356} (1995) 595.


\bibitem{eta_etaprime_pi} G. Beladidze \others,
  Phys. Lett. \textbf{B313}, (1993) 276.

\bibitem{eta_pi} A. Abele \others,
  Phys. Lett. \textbf{B423}, (1998) 175;
  D. Thompson \others,
  Phys. Rev. Lett. \textbf{79}, (1997) 1630.

\bibitem{etaprime_pi} E. I. Ivanov \others,
  Phys. Rev. Lett. \textbf{86}, (2001) 3977.

\bibitem{adam} A. Szczepaniak, M. Swat, A. Dzierba, and S. Teige,
  Phys. Rev. Lett. \textbf{91}, (2003) 092002.

\bibitem{qnp_tobias} T. Schl\"uter, \textit{Proceedings of the Sixth
    International Conference on Quarks and Nuclear Physics (QNP2012)},
  Paris (2012), PoS(QNP2012)074 [arXiv:1207.1076].

\bibitem{eta-etaprime_mixing} A. Bramon, R. Escribano, and M. Scadron,
  Eur. Phys. J. \textbf{C7} (1999) 271;
  T. Feldmann, P. Kroll, and B. Stech,
  Phys. Rev. \textbf{D58} (1998) 114006;
  S. Okubo and K. Jagannathan,
  Phys. Rev. \textbf{D15} (1977) 177.

\bibitem{eta-etaprime_ratio} F. Close and H. Lipkin,
  Phys. Lett. \textbf{B196} (1987) 245.


\bibitem{confx_vincent} V. Mathieu, these proceedings.

\bibitem{thesis_sebastian} S. Neubert, ``First Amplitude Analysis of
  Resonant Structures in the 5-Pion Continuum at COMPASS'', PhD
  Thesis, TU M\"unchen, 2012.

\bibitem{omega} A. Kirk, ``The CERN OMEGA Spectrometer:
  25 Year of Physics'', CERN 97-2 (1997) 31.

\bibitem{wa102_pipi} D. Barberis \others,
  Phys. Lett. \textbf{B453} (1999) 316.

\bibitem{qnp_alex} A. Austregesilo and T. Schl\"uter,
  \textit{Proceedings of the Sixth International Conference on Quarks
    and Nuclear Physics (QNP2012)}, Paris (2012), PoS(QNP2012)098
  [arXiv:1207.0949].

\end{thebibliography}
\end{document}